\documentstyle[12pt]{article}
\begin{document}

\begin{center}

{\large \bf Evolution of perturbations in a domain wall cosmology} \\

\vspace{1cm}

J\'ulio C\'esar Fabris\footnote{e-mail:fabris@cce.ufes.br} and
S\'ergio Vitorino de Borba Gon\c{c}alves\footnote{e-mail:svbg@if.uff.br}\\ 
\vspace{0.4cm}
{\it Departamento de F\'{\i}sica, Universidade Federal do
Esp\'{\i}rito Santo,} \\
\vspace{0.1cm}
{\it Vit\'oria CEP 29060-900-Esp\'{\i}rito Santo. Brazil.} \\

\vspace{1cm}

\end{center}
\centerline{\bf Abstract}
A fluid of domain walls has an effective equation of state $p_w = - \frac{2}{3}\rho_w$.
This equation of state is qualitativelly in agreement with the supernova type Ia observations.
We exploit a cosmological model where the matter content is given by a dust fluid
and a domain wall fluid. The process of formation of galaxies is essentially preserved.
On the other hand, the behaviour of the density contrast in the ordinary fluid is
highly altered when domain walls begin to dominate the matter content of the Universe.
This domain wall phase occurs at relative recent era, and its possible consequences are
discussed, specially concerning the Sachs-Wolfe effect.
\vspace{0.5cm}
\newline
Pacs numbers: 04.20.Jb, 98.80.Cq
\vspace{1cm}

One of the most surprising observational results in cosmology in the end of this
century is due to the use of the type Ia Supernova as standard candles in the evaluation
of luminosity distance as function of the redshift $z$. The two groups consacrated to this
program \cite{riess,perlmu1} arrived at the same conclusion that the Universe is now in an accelerated phase. The inflationary paradigm, initially restricted to the very early Universe,
was transferred to the Universe today, with strinking consequences, for example, for the
age of the Universe and many other cosmological parameters \cite{primack,ruth}. The most
accepted results \cite{efsta} indicate that the value of the decelerating parameter today
is given by $q = - \frac{\ddot aa}{\dot a^2} \sim - 0.66$. If an Universe filled by a perfect
fluid is considered, with an equation of state $p = \alpha\rho$, the evolution of the scale
factor is given by $a \propto t^\frac{2}{3(1 + \alpha)}$ and the decelerating paremeter reads
$q = \frac{1 + 3\alpha}{2}$. In this case, $q \sim - 0.66$ implies $\alpha \sim - 0.77$.
Hence, the Universe today should be dominated by a fluid with negative pressure such that
the strong energy condition is violated.
\par
One of the main issues related to this observational results, is the nature of this negative
pressure fluid. The position of the first acoustic peak in the spectrum of
the anisotropy of cosmic microwave background radiation is related to the total density of
the Universe. In spite of the fact that there is not untill now undoubtfull observational
results indicating where precisely this first acoustic peak is located, the recent
data coming from BOOMERANG and MAXIMA projects indicates that the density of the Universe
is near the critical density \cite{Hu,Lange}. Hence, it can be assumed that the Universe is spatially or
nearly spatially flat. On the other hand, the clustered mass is responsable for
$0.3 \sim 0.4$ of the critical density. Consequently, from this data it is possible to conclude
that $0.7 \sim 0.6$ of the total matter of the Universe is a smooth component.
This smooth component is generally called dark energy.
\par
A fluid of negative pressure
violating the strong energy condition
does not cluster at large scale. In particular, a cosmological constant, which can
be represented by an equation of state $p = - \rho$, remains perfectly smooth, since
its density fluctuations are exactly zero. However, the above mentioned results
for the deceleration parameter suggest a fluid different from the cosmological constant.
A very popular model to describe this dark energy is the so-called "quintessence", a
scalar field with an appropiate potential term such that the effective equation of state evolves
from a typical radiation equation of state ($p = \frac{\rho}{3}$) to a negative equation
of state \cite{wang,perlmu2,caldwell,martin}. But, it is difficult from the avaliable data to exclude others possibilities.
\par
In this work, we will study a "domain wall" cosmological model.
Domain walls are topological defects that appears in phase transitions in the early Universe,
like others kind of topological defects such as monopoles, cosmic strings and textures. When the topology of the vacuum manifold
exhibits disconnected regions we are facing domain walls: this comes out, generally, from
a breaking of a discrete symmetry group in the underline field model. Domain walls are characterized by the
$\pi_0$ homotopy group \cite{villenkin} and its energy-momentum tensor is
given by $T_{\mu\nu} = diag(\rho, 0, \rho, \rho)$. Hence, the equation of state along
the $y$ and $z$ spatial components is $p = - \rho$. A network of domain
walls may be represented by an isotropic equation of state $p = - \frac{2}{3}\rho$.
\par
From now on, we will not concentrate on a fundamental description of a network of
domain walls but, instead, we will consider a phenomenological description of this
network. Hence, the main point which will interest us is that this network of domain
walls may be described by the above mentioned equation of state which is in the
allowed range of possible values determined
by the supernova results. There are
also claims that the range of allowed values for $\alpha$ may be much more narrow \cite{martin}. However, it
seems that, in the present state of art, it is not possible to exclude a network of domain
walls as one of the possible realizations of the dark energy.
\par
In reference \cite{spergel} a domain wall dominated Universe has been studied using the
so-called solid dark matter model (SDM). In that work, the implications for the anisotropy of
cosmic background radiation has been addressed, using some modifications of the
CMBFAST code. Some specific signs of such SDM model have being identified.
In the present work, we return back to this problem, but trying to develop as far
as possible an analytical model. In this way, we intend to identify untill which extent
a domain wall phase will affect some observable quantities. Our goal is to control explicitly
some physical inputs concerning this domain wall phase.
\par
We will couple a domain wall with a pressurelless fluid. Hence, the effective equation of state
of this two fluid models evolves from $\alpha = 0$ to $\alpha = - \frac{2}{3}$. In this sense,
this phenomenological model exhibits some similarities with the quintessence model. However, the
fact that we are exploiting a hydrodynamical description will allow us to find exact solutions.
Moreover, we will perform a perturbative study of such model. In such perturbative study, we
will allow both fluids to fluctuate (what seems to be the correct way to treat the problem).
Then, the consequence of the existence of the domain wall fluid for the evolution of density
perturbations will be analysed. In fact, in principle the consequence of the existence of this
negative pressure fluid is profound, even with respect to the evolution of density perturbations
in the "ordinary" fluid: density contrast in the ordinary fluid stops to increase when
the domain wall fluid begins to determine the evolution of the Universe. But, comparison with some observational data shows that actually
the implications of the existence of that "exotic" fluid for the formation of structure and
the anisotropy of the cosmic microwave background radiation deserves a much more detailed analysis due to the fact that this fluid dominate the matter content of the Universe quite
recently.
\par
We consider a two-fluid model where
besides this domain wall component there is also dust whose equation of state is $p_d = 0$.
A dust-dominated phase must have
ocurred prior to the accelerated phase in order to allow gravitational instability
to generate local structures.
Hence, the field equations are
\begin{eqnarray}
\label{fe1}
R_{\mu\nu} - \frac{1}{2}g_{\mu\nu}R &=& 8\pi G\biggr[{\stackrel{d}{T}}_{\mu\nu} +
{\stackrel{w}{T}}_{\mu\nu}\biggl] \quad , \\
\label{fe2}
{\stackrel{d}{T}}^{\mu\nu}_{;\mu} = 0 \quad &,& \quad {\stackrel{w}{T}}^{\mu\nu}_{;\mu} = 0 \quad ,
\end{eqnarray}
where ${\stackrel{d}{T}}_{\mu\nu} = \rho_d u_\mu u_\nu$ and
${\stackrel{w}{T}}_{\mu\nu} = (\rho_w + p_w)u_\mu u_\nu - p_wg_{\mu\nu}$ are the energy-momentum tensor
for the dust and domain wall fluids, respectivelly.
\par
An isotropic and homogenous Universe is represented by the Friedmann-Robertson-Walker metric:
\begin{equation}
ds^2 = dt^2 - a^2(t)\biggr[\frac{dr^2}{1 - \epsilon r^2} + r^2(d\theta^2 + \sin^2\theta d\phi^2)\biggl] \quad ,
\end{equation}
where $\epsilon = 0, + 1, - 1$ describe a flat, closed and open spatial section respectivelly.
The corresponding equations of motion are
\begin{eqnarray}
\label{em1}
\biggr(\frac{\dot a}{a}\biggl)^2 + \frac{\epsilon}{a^2} &=& \frac{8\pi G}{3}\biggr(\rho_d
+ \rho_w\biggl) \quad , \\
\label{em2}
\dot\rho_d + 3 \frac{\dot a}{a}\rho_d = 0 \quad &,& \quad \dot\rho_w + \frac{\dot a}{a}\rho_w = 0
\end{eqnarray}
Since (\ref{em2}) imply $\rho_d = \frac{\rho_{d0}}{a^3}$ and $\rho_w = \frac{\rho_{w0}}{a}$,
there is just one equation to be solved:
\begin{equation}
\biggr(\frac{a'}{a}\biggl)^2 = c_1a + c_2a^3 \quad ,
\end{equation}
where $c_1 = \frac{8\pi G\rho_{d0}}{3}$, $c_2 = \frac{8\pi G\rho_{w0}}{3}$ and the primes
mean derivatives with respect to the conformal time $\eta$ defined as $dt = ad\eta$.
We have fixed also $\epsilon = 0$ (since this value is favoured by observation).
Defining $\sinh\theta = \sqrt{\frac{c_1}{c_2}}a$, this non-linear differential equation
reduces to
\begin{equation}
\int \frac{d\theta}{\sqrt{\sinh\theta}} = (c_1c_2)^{1/4}\eta \quad .
\end{equation}
The final solution for the scale factor is given as
\begin{equation}
a = \sqrt{\frac{c_1}{c_2}}\tan^2\biggl(\frac{x(\eta)}{2}\biggr) \quad , \quad x(\eta) = \mbox{am}\biggr((c_1c_2)^{1/4}\eta\biggl) \quad ,
\end{equation}
where $\mbox{am}(z)$ is the Jacobi amplitude function.
\par
The scale factor has two asymptotic regime. For small values of the cosmic time $t \rightarrow 0$,
$a \propto t^{2/3} \propto \eta^2$ ($0 < \eta < \infty$); for lage values of the cosmic time, $t \rightarrow \infty$,
$a \propto t^2 \propto \frac{1}{\eta^2}$ ($- \infty < \eta < 0$).
The effective equation of state evolves from $p_{eff} \sim 0$ to $p_{eff} \sim - \frac{2}{3}\rho_{eff}$, where $\rho_{eff} = \rho_d + \rho_w$. In fact, the deceleration paremeter evolves from $\frac{1}{2}$ to
$- \frac{1}{2}$. There is an initial dust dominated phase
followed by a domain wall dominated phase.
\par
Since the model developed above display a superluminal expansion for
large values of the cosmic time $t$, an important question is how fluctuations in the
ordinary and in the exotic fluid behave. Fluctuations in the dust fluid are affected in
two ways: first by the fact that the scale factor changes its behaviour; second, by the
fact that fluctuations in the dust are coupled to fluctuations in the domain wall fluid.
Since this two-fluid model is coupled through geometry, we will allow all fluids to fluctuate.
\par
As usual, let us introduce small fluctuations around the background solutions found before.
In the equations (\ref{fe1},\ref{fe2}) it is introduced the quantities
$\tilde g_{\mu\nu} = g_{\mu\nu} + h_{\mu\nu}$, $\tilde\rho_d = \rho_d + \delta\rho_d$,
$\tilde\rho_w = \rho_w + \delta\rho_w$, $\tilde u^\mu_d = u^\mu_d + \delta u^\mu_d$
and $\tilde u^\mu_w = u^\mu_w + \delta u^\mu_w$, where in each of these expressions,
the right-hand side represents a sum of the background solution and a fluctuation around it.
Note that the four-velocity for each fluid may fluctuate independently.
\par
The derivation of the equations which determine the evolution of these quantities is
standard \cite{weinberg}. We choose to fix the synchronous coordinate condition $h_{\mu0} = 0$. In this case,
the final equations are, in the conformal time coordinate,
\begin{eqnarray}
\label{pe1}
h'' + \frac{a'}{a}h' &=& - \frac{3}{2}\biggr[2\frac{a''}{a} -
3\biggr(\frac{a'}{a}\biggl)^2\biggl]\Delta_d - \frac{3}{2}\biggr[2\frac{a''}{a} -
\biggr(\frac{a'}{a}\biggl)^2\biggl]\Delta_w \quad ,\\
\label{pe2}
\Delta_d' + \Psi - \frac{h'}{2} &=& 0 \quad , \\
\label{pe3}
\Delta_w' + \Theta - \frac{h'}{2} &=& 0 \quad ,\\
\label{pe4}
\Psi' + \frac{a'}{a}\Psi &=& 0 \quad ,\\
\label{pe5}
\Theta' + 3\frac{a'}{a}\Theta + 2n^2\Delta_w &=& 0\quad .
\end{eqnarray}
In deriving these equations we have made the following redefinitions:
$h = \frac{h_{kk}}{a^2}$, $\Delta_d = \frac{\delta\rho_d}{\rho_d}$,
$\Delta_w = \frac{\delta\rho_w}{\rho_w}$, $\Psi = a{\delta u^i_d}_{,i}$,
$\Theta = a{\delta u^i_w}_{,i}$. Moreover, the spatial dependence of each quantity
is such that the Helmhotz equations is obeyed: $\nabla^2 Q(\vec x, t) = - n^2Q(\vec x,t)$
\par
Equations (\ref{pe1},\ref{pe2},\ref{pe3},\ref{pe4},\ref{pe5}) seems to admit no exact
solution, not only because of the coupling between all quantities, but mainly because of
the complicated form of the background expression for the scale factor.
However, it is possible to obtain analytical solutions in the asymptotic regimes.
Before to determine these analytical solutions, we remark that it is possible to
set $\Psi = 0$, since this quantity decouples from the others, and it contribute only
through a decreasing inhomogenous term. Hence, we can reduce the above system of equations
to just two coupled equations:
\begin{eqnarray}
\label{fs1}
\Delta_d'' + 3\frac{a'}{a}\Delta'_d &=& 3\Delta_w'' + 9\frac{a'}{a}\Delta_w' - 2n^2\Delta_w
\quad,\\
\label{fs2}
\Delta_d'' + \frac{a'}{a}\Delta_d' + \frac{3}{4}\biggr[2\frac{a''}{a} - 3\biggr(\frac{a'}{a}\biggl)^2\biggl]\Delta_d &=& - \frac{3}{4}\biggr[2\frac{a''}{a} -
\biggr(\frac{a'}{a}\biggl)^2\biggl]\Delta_w \quad .
\end{eqnarray}
\par
First, let us solve the perturbed equations for the dust phase. In this case, $a \propto \eta^2$.
and equation (\ref{fs2}) reduces to
\begin{equation}
\Delta_d'' + 2\frac{\Delta_d'}{\eta} - 6\frac{\Delta_d}{\eta^2} = 0
\end{equation}
leading to the well known solution for the evolution of the density contrast in a pure
dust Universe: $\Delta_d \propto \eta^2$. Hence, in the begining the domain wall fluid
do not influence either the background and the perturbed quantities. We can solve also the
homogenous equation for the fluctuation in the domain wall fluid, obtaining
\begin{equation}
\Delta_w = \eta^{-5/2}\biggr\{C_1I_{5/2}(\sqrt{\frac{2}{3}}n\eta) +
C_2K_{5/2}(\sqrt{\frac{2}{3}}n\eta)\biggl\} + C_3
\end{equation}
where $K_\nu(x)$ and $I_\nu(x)$ are the modified Bessel's functions and the $C_i$ are
integration constants.
\par
In the other asymptotic, the domain wall fluid dominate and the scale factor
behaves as $a \propto \eta^{-2}$, where $t \rightarrow \infty$ means $\eta \rightarrow 0_-$.
The coupled system (\ref{fs1},\ref{fs2}) reduces to
\begin{eqnarray}
\Delta''_d - 6\frac{\Delta_d}{\eta} &=& 3\Delta_w'' - 18\frac{\Delta_w'}{\eta}
- 2n^2\Delta_w \quad , \\
\Delta_d'' - 2\frac{\Delta_d'}{\eta} &=& - 6\frac{\Delta_c}{\eta^2} \quad ,
\end{eqnarray}
which can be expressed in terms of a single third order equation:
\begin{equation}
\Delta_w''' - 7\frac{\Delta_w}{\eta} + \biggr[-\frac{2}{3}n^2 + \frac{14}{\eta^2}\biggl]\Delta_w'
+\biggr[\frac{2}{3}\frac{n^2}{\eta} - \frac{14}{\eta^3}\biggl]\Delta_w = 0 \quad .
\end{equation}
This equation can be solved remembering that the synchrounous coordinate condition
has a residual coordinate freedom \cite{peebles}. Using this fact, it is easy to see that $\Delta_w \propto \eta$
is a solution of the third order differential equation. Hence,
it is possible to reduce the order of the equation obtaining the final solution
\begin{equation}
\Delta_w = \eta\biggr\{\int\eta^\frac{5}{2}\biggr[D_1I_{5/2}(\sqrt{\frac{2}{3}}n\eta) +
D_2K_{5/2}(\sqrt{\frac{2}{3}}n\eta)\biggl] + D_3\biggl\} \quad ,
\end{equation}
where the $D_i$ are again integration constants.
\par
We are now able to analyse the results obtained.
The first important point is that in the begining of the dust dominated phase,
the domain wall fluid plays no important role: the dust density contrast evolves
exactly as in a pure dust model. Since, it is at this period that galaxies form,
it is possible to state that the presence of the domain wall fluid will not spoil
this scenario. The density contrast for the domain wall fluid exhibits, on the other
hand, features that deserves some comments. In the long wavelength limit $n \rightarrow 0$,
it presents a constant mode and a decreasing mode. However, in the small wavelength limit
$n \rightarrow \infty$, it displays an exponential increasing mode. This may lead to instabilities.
However, it has been shown in \cite{jerome,sergio} that these instabilities are consequence of
the hydrodynamical approach employed here. The hydrodynamical approach is a phenomenological
description, which mimics some features of this fluid of topological defects.
However, when we substitute this hydrodynamical approach by a field description, the
instabilities in the small wavelength limit disappears, keeping the behaviour in
the long wavelength limit unaltered. It has been showed in \cite{sergio}
that the long wavelength limit is insensitive to the approach used.
\par
In the other asymptotic regime, the domain wall fluid exhibits only decreasing modes.
In fact, when $n \rightarrow 0$, we find
\begin{equation}
\Delta_w \sim D_1\eta^6 + D_2\eta^2 + D_3\eta
\end{equation}
which goes to zero as time evolves.
In this same limit, the density contrast in the dust fluid exhibits decreasing and constant modes:
\begin{equation}
\Delta_d \sim D_1\eta^7 + D_2\eta^2 + D_3 \quad .
\end{equation}
Hence, as the domain wall fluid dominates the matter content of the Universe,
density perturbations in the ordinary fluid do not grow anymore. This is essentially
due to the coupling of both fluids at perturbative level.
\par
The main question that comes out from these results is if there is other observational
consequences. In fact, this is a much more difficult question for the following reasons.
Observations today indicate that around $40\%$ of the matter of the Universe suffers
the process of gravitational collapse, while the others $60\%$ remain a smooth component.
Accepting that the Universe is flat, we have than
$\Omega_c \sim 0.4$ and $\Omega_s \sim 0.6$ today, where the subscripts designates the
clustered and the smooth components of the Universe. The clustered component
may be not baryonic but it is quite possibly a cold component, i.e., a pressurelless fluid,
like the ordinary matter introduced in the model developed above. The smooth component
must violate the strong energy condition, as the domain wall fluid considered here.
Since $\Omega_c \propto a^{-3}$ and $\Omega_s \propto a^{-1}$, fixing the value of the
scale factor equal to one today, we can evaluate when the smooth component begins to dominate
the matter content of the Universe. Imposing $\Omega_c = \Omega_s$ and remembering that
the redshift is given by $z = - 1 + \frac{a_0}{a}$, we find that the domain wall fluid
dominated era begins at $z \sim 0.22$. Accepting that the age of the Universe is
$t_0 \sim 13\,Gy$, this happens at $t \sim 11.7\,Gy$. This is quite recent.
\par
Of course, this domination of the domain wall fluid drives an accelerated expansion, which
is reflected in the high redshift supernova measurements. For the anisotropy of the cosmic
microwave background radiation the situation is much less clear. If we think on the
integrated Sachs Wolfe effect, the domain wall dominate at about $10\%$ of the integration
interval. But we must be cautious in saying that this would lead to no modification at
all with respect to a pure dust Universe since we must verify how different multipole moments
are affected during the travel from the last scattering surface to the observer today.
It may happen that for some class of multipole moments this modification at the very end
of the trajectory leads to a quite different behaviour.
\par
However, even qualitativelly it is possible to verify that the domain wall dominated phase will
have some consequences for the anisotropy of the cosmic microwave background radiation.
This anisotropy is determined by the Sachs-Wolfe formula
\begin{equation}
\frac{\Delta T}{T} = \frac{1}{2}\int_{\eta_e}^{\eta_r}\frac{d}{d\eta}\biggr(\frac{h_{ij}}{a^2}\biggl)e^ie^jd\eta \quad ,
\end{equation}
where we have tried to keep our previous definitions; $\eta_e$ and $\eta_r$ are the emission
and reception time and $e^i$ is an unitary vector defining the direction of
observation.
The perturbed metric $h_{ij}$ may be decomposed conveniently as
\begin{equation}
h_{ij} = a^2\biggr(h_1\delta_{ij}Q + \frac{h_2}{n^2}Q_{,i,j}\biggl) \quad .
\end{equation}
Hence, $h = 3h_1 + h_2$. It is possible to writte down equations governing the behaviour
of these metric functions \cite{grisha}. Transposing these equations for our definitions,
we obtain in particular
\begin{equation}
h_1'' + 2\frac{a'}{a}h_1' = -\biggr[2\frac{a''}{a} - \biggr(\frac{a'}{a}\biggl)^2\biggl]\Delta_w \quad.
\end{equation}
In the long wavelength limit this equation leads to
\begin{equation}
h_1' \sim D_1\eta^5 + D_2\eta + D_3 + D_4\eta^4
\quad,
\end{equation}
with a similar expression for $h_2'$. Hence, a constant mode will appear in the integral
of the Sachs-Wolfe effect. In spite of the fact that density perturbations remain constant
or decrease, we must expect important distortions in the cosmic microwave background radiation.
It is important to note that the constant mode is associated to the residual coordinate freedom:
but this mode may have physical meaning due to the junction conditions between the different
phases of evolution of the Universe.
\par
Domain walls are topological defects, metastable, which could
be originated in phase transitions in the primordial Universe. It has already be argued that
domain walls may play an important role in the evolution of the Universe \cite{villenkin}. Indeed, domain walls
may lead to an accelerated Universe, with a value for the decelerating parameter compatible with
the results coming from supernova type Ia observations.
We have verified here that the presence of this fluid does not spoil the formation of galaxies
process at the begining of the dust dominated phase. However, in the deep domain wall phase,
matter perturbations do not grow anymore due to the coupling with the domain wall fluid.
Even if the domain wall fluid would dominate for a small interval of time, this may have
important consequences.
\par
The goal of the present paper was to develop an analytical domain wall cosmological model, verifying in
particular its
consequence for the evolution of density perturbations and for
the Sachs-Wolfe effect. In a preceding work \cite{sergio1},
we have studied a two-fluid model where ordinary matter was coupled to a cosmic string
fluid. A network of cosmic string may be represented by an equation of state
$p = - \frac{\rho}{3}$. Hence, in principle such fluid is outside the observational
limits imposed by supernova type Ia observations, in opposition to what happens in
the case of domain walls. But, there are other important differences: when the string
fluid dominates the evolution of the Universe, perturbations in the ordinary fluid may
still grow, although very slowly. In the case of domain walls, perturbations
in the ordinary fluid simply ceases to grow. In some sense, this is due to the fact that
in the case of cosmic string, density contrast in the cosmic string fluid decouples from
density contrast in the ordinary fluid, what does not happen in the case of domain walls.
As it has been argued before, this may lead also to consequences in what concerns the
anisotropy of cosmic microwave background. We intend to perform such analysis from
the fundamental definition of the Sachs-Wolfe effect in terms of the metric functions.
\vspace{0.5cm}
\newline
{\bf Acknowledgements}: We thank CNPQ (Brazil) for partial financial support.

\end{document}